\documentclass[11pt,twoside]{article}

\usepackage{asp2006}
\usepackage{graphicx}

\begin{document}

\title{On the Connection between Mass Loss and Evolution of C-rich 
AGB stars}   

\author{Lars Mattsson,$^1$ 
Susanne H\"ofner,$^1$ 
Rurik Wahlin,$^1$ and 
Falk Herwig,$^2$ }  

\affil{$^1$Dept. of Astronomy \& Space Physics, Uppsala University, 
Uppsala, \hspace*{0.3cm}Sweden  \\   
$^2$Theoretical Astrophysics Group, LANL, Los Alamos NM, USA}  
 
\begin{abstract} 
The mass-loss properties of carbon AGB stars are not very well 
constrained at present. A variety of empirical or theoretical 
formulae with different parameterisations are available in the 
literature and the agreement between them is anything but good. 
These simple mass-loss prescriptions are nonetheless used in many 
models of stellar evolution without much consideration of their 
applicability in various cases.  We present here an on-going project 
aiming at a better description of the mass loss, that could be used 
to improve stellar evolution models -- especially the evolution 
during the TP-AGB phase. As a first step, we have considered the 
mass-loss evolution during a He-shell flash. Using stellar 
parameters sampled from a stellar evolutionary track, we have 
computed the time evolution of the atmospheric layers and wind
acceleration region during a flash event with detailed 
frequency-dependent radiation-hydrodynamical models including dust 
formation. We find that existing simple mass-loss prescriptions 
imply mass-loss evolutions different than our model. Based on these 
results, we have also simulated the subsequent long-term dynamical 
evolution of the circumstellar envelope (CSE), including the 
formation of a detached shell. The second step of the project deals 
with the dependence of mass loss on the basic stellar parameters. 
At the moment we are computing a large grid of wind models for 
C-rich AGB stars. Preliminary results show that simple 
parameterisations are difficult to obtain in certain regions of 
the parameter space considered, due to strong non-linearities in 
the wind mechanism.
\end{abstract}

\section{Introduction} 
Models of mass loss on the AGB are essential for understanding the 
late stages of stellar evolution. In practice, the modelling of 
stellar evolution requires a prescription of mass-loss rates as a 
function of basic stellar parameters. Empirical studies, such as 
the classical one by Vassiliadis \& Wood (1993), can only provide 
a limited amount of information, since one cannot easily study the 
effect of individual parameters. The reason for this is of course 
that the evolution of the stellar parameters is interconnected and 
we can only observe stars at a given evolutionary stage. 
Observations can tell us about the mass loss for a given set of 
stellar parameters, but the exact dependences on these parameters 
are unfortunately degenerated. 

Numerical modelling has one major advantage here: it can be used as 
a tool for constraining the actual physics involved in wind 
formation. From the early efforts, e.g.\ by Bowen (1988) and Fleischer 
et al.~(1992), to our latest models \citep{Hofner03, Mattsson07}, 
a lot of the physics behind dust-driven winds has been brought to 
light. Much work still remains to be done, but we have now reached a 
reasonable level of realism in the modelling, as shown by detailed
comparison with observations. A picture has emerged with dust-driven 
winds as a complex phenomenon, where the onset of mass loss requires 
particular circumstances. In fact, C stars with strong dust-driven 
winds may be a rare species even if most C stars apparenly lose mass. 
With this in mind, one may ask: What are the consequences for stellar 
evolution, nucleosynthesis, and consequently the chemical evolution 
of galaxies?  

\section{Mass-Loss Modulations during the TP-AGB Phase}

The common hypothesis about how detached shells around TP-AGB stars 
are formed is that these structures are a consequence of the mass-loss 
modulations during the phase of He-shell flashes/thermal pulses 
\citep{Olofsson90}. Previous modelling has shown that interactions 
between different wind phases that arise from variations in the wind 
velocity and mass-loss rate are capable of producing such shells 
\citep{Steffen00}. 

  \begin{figure}
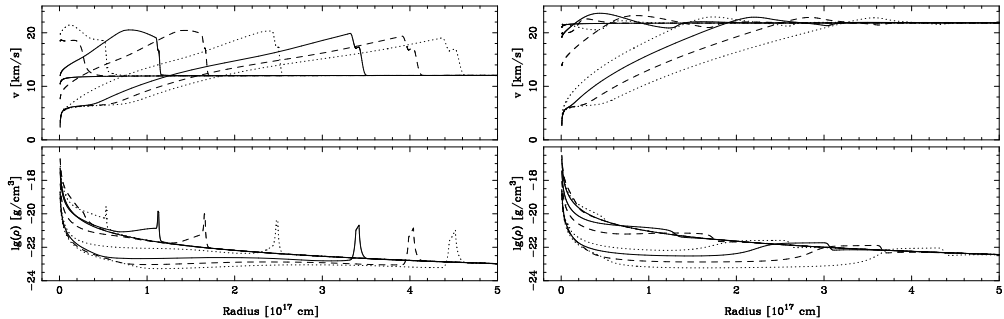

  \resizebox{\hsize}{!}{
  \includegraphics[width=5cm, angle=270]{mattsson_fig1a}
  \includegraphics[width=5cm, angle=270]{mattsson_fig1b}}
  \caption{  \label{fixedtemp}
  Formation and dynamical evolution of a detached shell: snapshots 
  of the velocity and density structure according to the CSE model. 
  {\it Upper panels\/}: the velocity field; {\it lower panels\/}: 
  the logarithmic density. {\it Left\/}: piston amplitude 
  $\Delta v_{\rm p}=4.0$ km~s$^{-1}$; {\it right\/}: piston 
  amplitude $\Delta v_{\rm p}=6.0$ km~s$^{-1}$.
  }
  \end{figure}

Using the RHD code for dynamic atmospheres by H\"ofner et al.\ (2003) 
in combination with a stellar evolutionary track of a $M = 2$ M$_\odot$, 
$Z = 0.01$ star as input \cite[see][for details about the stellar 
evolution model]{Herwig00, Herwig04}, we have calculated the wind 
evolution during a He-shell flash \citep{Mattsson07}. With this 
evolution computed, we constructed a time-dependent inner boundary 
condition for a hydrodynamic model of the circumstellar envelope (CSE) 
evolution. The observed wind and mass-loss properties associated with 
detached shells \cite[e.g.][]{Olofsson00, Schoier05} 
are rather well reproduced using stellar parameters sampled from the 
evolutionary track. We tried different amplitudes for the piston 
boundary condition which is used to simulate the stellar pulsation 
and found that too high amplitudes lead to very little wind interaction 
and thus no shell structure is formed (see Figure \ref{fixedtemp}). 
Too low amplitudes, on the other hand, tend to prohibit any wind 
formation at all. It seems that for the stellar parameters investigated 
here, the optimal piston amplitude is $\sim 4$ km~s$^{-1}$, and this 
value may perhaps be used to constrain the pulsation properties.

It is also interesting to note that existing mass-loss prescriptions 
for AGB stars imply mass-loss evolutions that are not only different 
in comparison with our detailed model, but are rarely consistent with 
each other, either \cite[see][for further discussion]{Mattsson07}. 
Furthermore, constraints on the mass-loss evolution derived from 
observations \citep{Olofsson00, Schoier05} appear to agree better 
with our results than with these mass-loss formulae. 

\section{Mass Loss as a Function of Stellar Parameters}   

We are currently in the process of computing a large grid of dynamical 
C star atmosphere models.  The overall purpose of the grid is to 
explore how the mass-loss rate depends on the stellar parameters. 
In a future step of this project we will consider how pulsations and 
winds affect the spectral properties of C stars as well, but here
we will discuss only the wind formation and the mass loss itself.

\begin{figure}[!ht]
  \includegraphics[width=6.5cm]{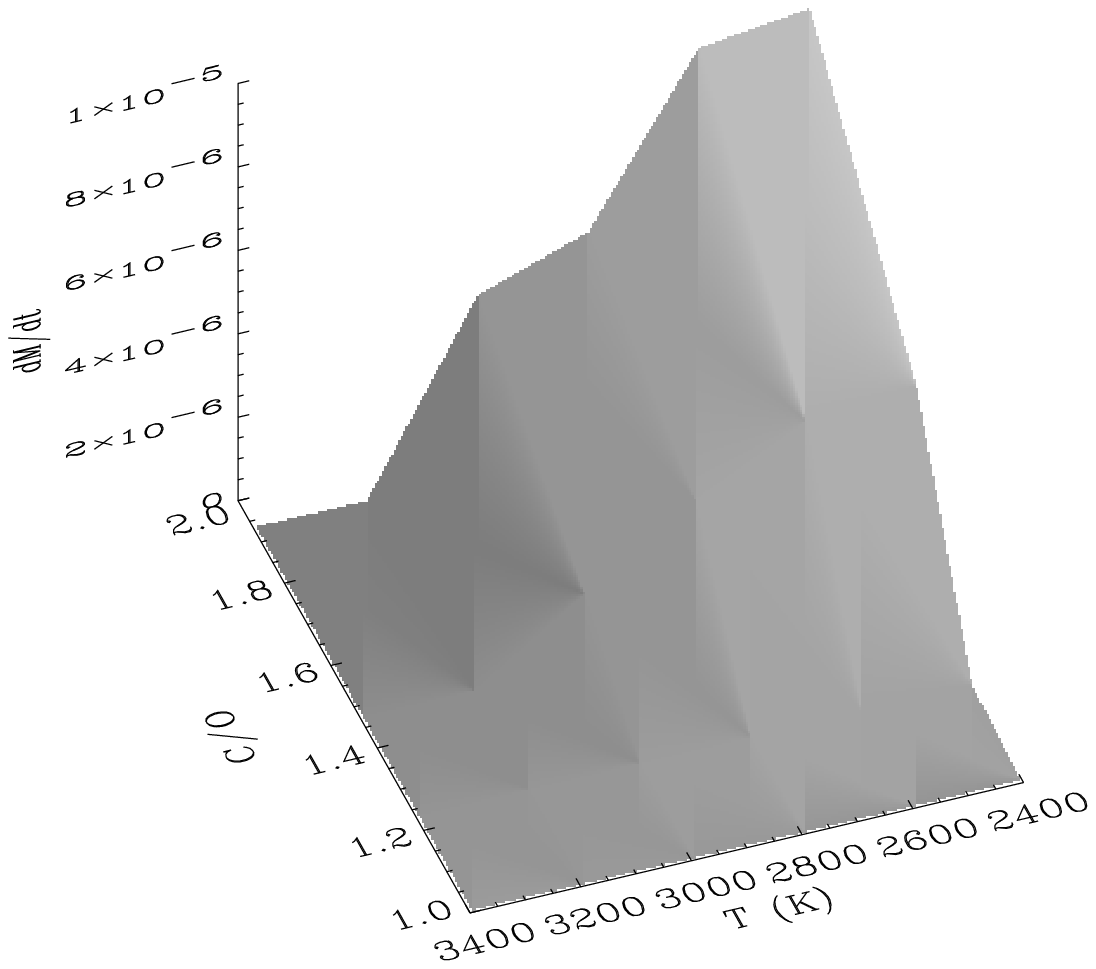}
  \includegraphics[width=6.5cm]{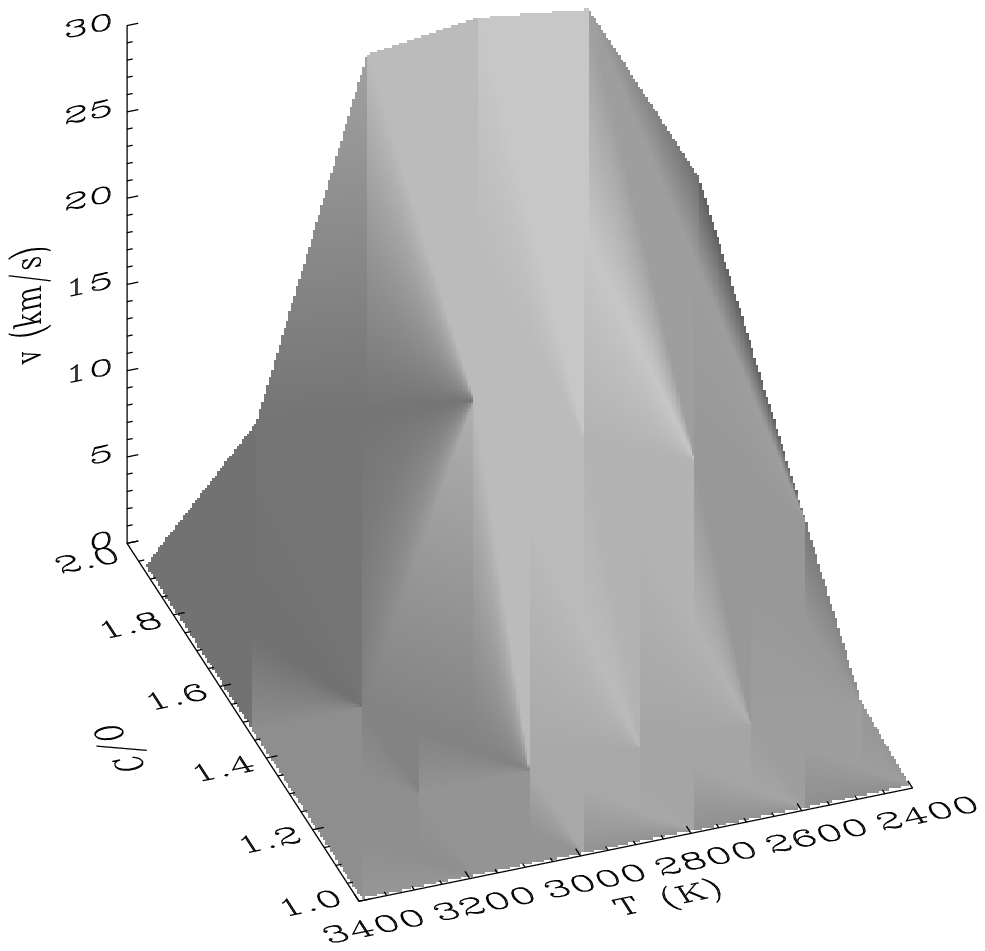}
  \caption{Mass loss in M$_\odot$ yr$^{-1}$ ({\it left\/}) and wind 
  velocity ({\it right\/}) as functions of $T_{\rm eff}$ and ${\rm C/O}$ 
  for a C star with $L_\star = 10^4$ L$_\odot$, $M = 1$ M$_\odot$, 
  $Z =$ Z$_\odot$ and $\Delta v_{\rm p}=4.0$ km~s$^{-1}$, 
  according to our detailed numerical wind model.
  \label{trends}
  }
  \end{figure}

Before summarising the numerical results, it is worthwhile to consider 
some fundamental aspects of the onset of dust-driven winds from a basic 
physics point of view. The condensation of dust grains is very 
sensitive to temperature and is prohibited by high gas temperatures 
and low densities, which means that every C star has a {\it condensation 
radius} ($R_{\rm c}$) -- a critical radius inside which dust grains 
cannot condense -- as well as an outer radius where the condensation 
stops.  Furthermore, in order to levitate the material out to 
$R_{\rm c}$, the acceleration of the gas due to pulsations and 
radiation must be large enough. There cannot be any circumstellar dust 
formation and hence no dust-driven wind if insufficient material is 
transported out to $R_{\rm c}$. The location of $R_{\rm c}$ will depend 
on the effective temperature of the star, as well as on its photospheric 
radius. Then, quite obviously, there must be a temperture-dependent 
threshold for the formation dust-driven winds. The sudden drop in the 
mass-loss rate and wind velocity above $T_{\rm eff}\approx 3000$ K 
is probably even steeper than Figure \ref{trends} seems to indicate, 
due to a grid spacing of $200$ K. 

Another very obvious threshold emerges as we consider the wind 
acceleration process after the onset of dust formation.  Beyond the 
condensation radius $R_{\rm c}$, the criterion for wind formation is 
that the radiative acceleration must be strong enough to overcome the 
gravitational field of the star, i.e. $\Gamma_{\rm dust} > 
\Gamma_{\rm crit}$, where $\Gamma$ is the ratio of the radiative to 
the gravitational acceleration \cite[see e.g.][]{Dominik90}. The 
radiative acceleration is proportional to the dust opacity 
$\kappa_{\rm dust}$, which is a function of the abundance of free 
carbon $\tilde{\varepsilon}_{\rm C}$ and the mean degree of dust 
condensation $f_{\rm c}$ \cite[cf.][]{Hofner97}. Since 
$\kappa_{\rm dust}\propto {\rho_{\rm dust}\over\rho_{\rm gas}}$, 
we have that
\begin{equation}
\Gamma_{\rm dust} \propto {\rho_{\rm dust} \over \rho_{\rm gas}} 
{L_\star \over M_\star},
\quad
{\rho_{\rm dust} \over \rho_{\rm gas}} = {m_{\rm C} \over 
m_{\rm H} + m_{\rm He} \varepsilon_{\rm He}}
\,\tilde{\varepsilon}_{\rm C}f_{\rm c} = 
{12\over 1.4}\left({\varepsilon_{\rm C}\over\varepsilon_{\rm O}}
-1\right)\,\varepsilon_{\rm O}f_{\rm c}.
\end{equation}
Thus, for a given mass and luminosity, there must exist a critical 
value of ${\rm C/O}$ (or more exactly 
$\varepsilon_{\rm C}-\varepsilon_{\rm O}$) 
for which $\Gamma_{\rm dust} \equiv \Gamma_{\rm crit}$. 
We also note that this critical ${\rm C/O}$ cannot be arbitrarily 
close to unity for realistic stellar parameters.

Let us now turn to the numerical modelling results. We present here a 
sub-grid of models of stars with $L_\star = 10^4$ L$_\odot$, 
$M = 1$ M$_\odot$, $Z =$ Z$_\odot$ and $\Delta v_{\rm p}=4.0$ 
km~s$^{-1}$, where we vary $T_{\rm eff}$ and ${\rm C/O}$. 
To eliminate the pulsation period (free parameter), we employ an 
empirical period-luminosity relation \citep{Feast89}. The predicted 
threshold appears as expected and we find that for ${\rm C/O} < 1.2$ 
and/or $T_{\rm eff} > 3200$ K no dust-driven wind is formed with the 
other parameters fixed as above (see Fig. \ref{trends}).
Thus, any mass loss in this part of parameter space is likely due to 
some other mechanism. Since the surface gravity scales with 
$T_{\rm eff}$, it is quite expected that mass loss scales with 
$T_{\rm eff}$, too. But we also see a rather strong dependence on 
${\rm C/O}$ for both the wind velocity and the mass-loss rate, which 
is quite interesting in comparison with previous studies of this kind. 
Arndt et al.\ (1997) as well as Wachter et al.\ (2002) argue for a 
weak dependence on ${\rm C/O}$, which stands in sharp contrast to the 
results presented here. This is probably due to the parameter range for
which these formulae where developed, as well as differences in the 
basic model assumptions.

\section{Effects on Stellar Evolution and Nucleosynthesis}

Mass loss affects stellar evolution is various ways. For C stars the 
duration of the AGB phase and the number of thermal pulses is almost 
solely determined by the mass-loss rate. The internal structure 
depends on the mass-loss rate as well \citep{Blocker95}, which in turn 
affects the fundamental stellar parameters. Since the mass-loss rate 
depends on these stellar parameters, there is an interesting feedback 
mechanism at work here, which means that the mass-loss prescription 
put into a stellar evolution model is critical. The observed 
initial-final mass relation is our best empirical constraint, but it 
can just tell us the total mass lost during the RGB and AGB phase. 
Hence, much of the AGB evolution will only be guess-work as long as 
we do not understand mass loss properly. 

Perhaps even more interesting, in the context of this conference, is 
the effect on nucleosynthesis. It is well known that varying the 
mass-loss rate can have profound effects on the chemical yields for 
AGB stars \cite[see e.g.][]{Hoek97}.
As we pointed out above, the mass-loss rate determines the number of 
thermal pulses, which represents the maximum possible number of third 
dredge-up (TDU) episodes. Furthermore, it can also determine if and 
when the TDU begins and ends \citep{Karakas02}. Needless to say, a 
correct implementation of mass loss is essential to obtain realistic
yields for AGB stars, and the AGB contribution to the chemical 
evolution of galaxies therefore still has to be regarded as uncertain. 
This also becomes evident when different sets of AGB yields are 
compared \citep{Olofsson07}.

\section{Concluding Remarks}

Existing mass-loss prescriptions for AGB stars are inconsistent with 
each other and simple mass-loss prescriptions might be hard to obtain. 
Here we have tried to show that it may be dangerous to use parametric 
mass-loss formualae including too few stellar parameters. Moreover, 
the first results from our new detailed model grid suggests that C 
stars with strong winds may be quite rare. How would this affect 
stellar evolution? Nucleosynthesis? Chemical evolution? The present 
dynamic atmosphere models compare well with observations and it is 
therefore likely that we are beginning to understand the wind 
mechanism well enough to be able to reduce the current uncertainties 
related to the mass-loss evolution of C stars. 

\acknowledgements 
This work was partly supported by the Swedish Research Council 
(Vetenskapsr\aa det).

\question{Speck} You state that for C/O $<$ 1.2 mass-loss doesn't 
happen.  But the literature suggests that C/O $<$ 1.2 for {\it most} 
C stars.  And most C stars have dust (SiC 11~$\mu$m feature). 
How do you reconcile these?

\answer{Mattsson} Yes, about 75\% of all observed C stars have 
C/O ratios below 1.2, but still mass-loss is detected. What our models 
tell us is that below 1.2 there is no strong, dust-driven wind. The 
observed mass-loss must be due to some other mechanism, if the dust 
is there.

\question{Gallino} Maybe combining with $\geq1.5$ M$_{\odot}$  models
is a better choice. For 1 M$_{\odot}$ no TDU is typically found.

\answer{Mattsson} Yes, this is just a first result. The grid will be 
vastly extended to cover all relevant parameter ranges.

\question{Willson} You slid by the piston amplitude issue. Good piston 
constraints are essential for a good theoretical \.M relation. We found
the following very important: Keep track of the work done on it by the 
piston.  Limit this to be $< L$. See Cox \& Giuli. This gave 
$d\log\dot{M}/d\log L$ consistent with Vassiliadis \& Wood, the only 
reliable empirical determination.

\answer{Mattsson} Very well. If the pulsation were too strong we would 
see this in the luminosity variations, but we don't.

\question{Aringer} I also see a problem that there is a large number 
of carbon stars with higher temperature and lower C/O losing mass.
Maybe dust is not the driver there. However, do not trust any published 
C/O ratios for AGB variables. There are large uncertainties.

\end{document}